\documentclass[12pt,a4paper]{article}
\usepackage{color,graphics,amsmath,epsfig,rotating,cite}

\newcommand\beqn{\begin{eqnarray}}
\newcommand\eeqn{\end{eqnarray}}

\def\x{\chi}

\def\nt{\tilde \x^0}
\def\ch{\tilde \x^+}
\def\chpm{\tilde \x^\pm}

\def\micro{{\tt micrOMEGAs}}

\newcommand{\mnt}[1]   {m_{\tilde\x^0_{#1}}}

\begin{document}

\begin{flushright}
LAPTH-Conf-1161/06\\
CERN-PH-TH/2006-187
\end{flushright}
\vspace*{4mm}

\begin{center}

{\Large\bf
   Neutralino dark matter in the MSSM with CP violation}\\[6mm]

{\large
   G. B\'elanger$^1$, F. Boudjema$^1$,
   S. Kraml$^2$, A. Pukhov$^3$, A. Semenov$^4$}\\[6mm]

{\it
1) LAPTH, 9 Chemin de Bellevue, B.P. 110, F-74941 Annecy-le-Vieux , France
\footnote{UMR
5108 du CNRS associ\'ee \`a l'Iniversit\'e de Savoie}\\
2) CERN, Dept. of Physics, Theory Division, CH-1211 Geneva 23, Switzerland\\
3) Skobeltsyn Inst. of Nuclear Physics, Moscow State Univ., Moscow 119992, Russia\\
4) Joint Institute for Nuclear Research (JINR), 141980, Dubna, Russia}\\[6mm]

\end{center}

\begin{abstract}
In the MSSM, the presence of CP phases can have a significant
impact on the neutralino relic abundance. These phase effects
are on the one hand due to shifts in the masses, on the other hand
due to modifications of the couplings. Typical variations in
$\Omega h^2$ solely from modifications in the couplings are ${\cal
O}$(10\%--100\%).
\end{abstract}


\section{Introduction}

With the conclusive evidence for a significant component of cold
dark matter (CDM) in the Universe, there is considerable interest,
both at the theoretical and experimental level, to identify this
CDM and analyse its properties. In particular, if the CDM consists
of a new weakly interacting massive particle (WIMP), for example
the lightest neutralino in supersymmetric models, the next
generation of colliders has good prospects to discover it and
determine its  properties. If the parameters of the underlying
model can be measured with sufficient precision, it may be
possible to   predict  the annihilation cross sections  hence the
thermal relic density of the CDM candidate. The aim is to reach an
accuracy comparable to the one from cosmological measurements to
be able to perform consistency checks between a particular model
of new physics and cosmology ~\cite{Allanach:2004xn,Baltz:2006fm}.
The recent data from WMAP and SDSS imply
a value for  the relic density of cold dark matter
$\Omega h^2=0.105\pm 0.008$ ~\cite{Tegmark:2006az} at 2$\sigma$
    \footnote{In the figures we will rather use the 2$\sigma$ range as of 2003,
  $0.0945 <\Omega h^2 < 0.1287$
  ~\cite{Spergel:2003cb,Tegmark:2003ud} and refer to this as
the WMAP range. Note that the exact value and uncertainty of $\Omega h^2$ extracted from
cosmological data depend on both the precise data sets used and the assumptions about other
parameters~\cite{Lahav:2006qy}.}.

Most of the studies that have either analysed the constraints on
the parameter space of the MSSM arising from the relic density
measurement, or that have analysed how to extract precise
information on the model parameters at future colliders have
assumed that CP is conserved.
 However CP-violating phases are generic in the MSSM. Furthermore
 CP violation may help to generate
the correct baryon asymmetry in the Universe in scenarios of
electroweak baryogenesis ~\cite{Balazs:2004ae}.

CP phases can have a strong impact on   the relic density of
neutralino dark matter, both due to modifications in the sparticle
couplings and due to changes in the physical masses. Here we
highlight the impact of CP phases  in a few typical scenarios  for
which the LSP is a `good' CDM candidate ~\cite{Belanger:2006qa}.
Some of the largest effects are in fact due to kinematics. This
should be expected as the relic density is often very sensitive to
masses, in particular to the exact mass difference between the LSP
and NLSP in coannihilation processes, or in the case of
annihilation near a $s$-channel Higgs resonance, to the difference
between twice the LSP mass and the mass of the Higgs. In these
scenarios, setting apart the purely kinematic effects hence
somewhat tames the huge effects due to phases found in some of
the early studies~\cite{Gondolo:1999gu}. On the other hand, there
are other cases where the phase dependence of masses and
couplings work against each other. Taking out the kinematic
effects actually enhances the phase dependence of the number
density. Since what are relevant to experiments are rather the
physical observables (masses, branching ratios, etc.) than the
underlying parameters, we take special care to disentangle effects
arising from changes in the couplings from purely kinematic
effects.

\section{The model}

In the MSSM, the parameters that can have CP phases are the
gaugino and Higgsino mass parameters, $M_i=|M_i|e^{i \phi_i}$ and
$\mu=|\mu|e^{i \phi_\mu}$, and the trilinear sfermion-Higgs
couplings, $A_f=|A_f|e^{i \phi_f}$. For the relic density of the
LSP the relevant phases are those of the neutralino sector (only
$\phi_1$ and $\phi_\mu$ since $\phi_2$ can be rotated away) and
the phase of the trilinear coupling $A_t$ which affects the Higgs
sector. The parameters that will be allowed to vary are those of
the MSSM defined at the weak scale
\begin{equation}
  |M_1|,\; |\mu|,\; \tan\beta,\; m_{H^+},\; |A_t|,\; M_{S},\;
  \phi_1,\; \phi_\mu,\; \phi_t,\; \phi_l \,.
\end{equation}
where  universality at the GUT scale for the gaugino masses is
assumed. $M_S$ is the common mass for third generation sfermion,
the common mass for the  sfermions of the first and second
generation is set to $m_{\tilde f_{L,R}}=10$~TeV and all trilinear
couplings with the exception of $A_t$ are set to $|A_f|=1$~TeV.
 The phase of the selectron coupling
$\phi_e$, although irrelevant for the relic density has to be
taken into account since it contributes to electric dipole moments
(EDMs).

Allowing for CP-violating phases  induces a mixing between the two
CP-even states $h^0$, $H^0$ and the CP-odd state $A^0$ of the
MSSM.  The resulting mass eigenstates $h_1,h_2,h_3$ (
$m_{h_1}<m_{h_2}<m_{h_3}$) are no longer eigenstates of CP,
therefore it is preferable to use the charged Higgs mass,
$m_{H^+}$, as an independent parameter.  The Lagrangian for the
interaction of the lightest neutralino with Higgs bosons which
governs the neutralino annihilation cross section via Higgs
exchange, writes
\begin{equation}
  {\cal L}_{{\tilde \chi^0_1}{\tilde \chi^0_1} h_i}=-
  \frac{g}{2}{\sum_{i=1}^3}\,\overline{\tilde\chi_1^0} (
  g^{S}_{h_i{\tilde \chi^0_1}{\tilde \chi^0_1}}+ i \gamma_5
  g^{P}_{h_i{\tilde \chi^0_1}{\tilde \chi^0_1}} )\tilde{\chi}_1^0 h_i
\label{eq:gs}
\end{equation}
with  the scalar and pseudoscalar couplings corresponding to the
real and imaginary part of the same expression, see
~\cite{Belanger:2006qa}. These couplings depend on the neutralino
mixing, hence on phases in the neutralino sector,  as well as on
phases that enter the Higgs mixing, for example $\phi_t$.
 Indeed, in the MSSM, the
Higgs CP mixing is induced by loops involving top squarks and is
proportional to
$Im(A_t\mu)/(m^2_{{\tilde{t}}_2}-m^2_{{\tilde{t}}_1})$~\cite{Choi:2002zp}.
Thus a large mixing is expected when $Im(A_t\mu)$ is large as
compared to the square of the stop masses.

\section{Relic density of dark matter}

The results presented here have been obtained with the new
implementation of the CPV-MSSM  within
\micro2.0~\cite{Belanger:2006is}. In this code,  CP phases are
taken into account consistently in all annihilation and
coannihilation channels.  The computation of masses, mixings and
effective couplings in the Higgs sector relies on {\tt
CPsuperH}~\cite{Lee:2003nt}. Standard \micro~ routines are used to
calculate the relic density of dark matter
~\cite{Belanger:2004yn}.

The cross sections for the annihilation and coannihilation
processes will depend on phases, and so will the
thermally-averaged cross section ~\cite{Belanger:2006qa}. Part of
this is due to changes in the physical masses, leading to huge
variations in the relic density especially when coannihilation
processes are important or when annihilation occurs near a
resonance. We will therefore take special care to disentangle the
effects from kinematics and couplings.

At vanishing relative velocity, $v\to 0$, neutralino annnihilation
through $s$-channel scalar exchange is $p$-wave suppressed; the
annihilation proceeds strictly through pseudoscalar exchange.
Nevertheless when performing the thermal averaging, the scalar
exchange cannot be neglected altogether. In the MSSM with real
parameters it can amount to ${\cal O}(10\%)$ of the total
contribution. In the presence of phases, all the neutral Higgs
bosons can acquire a pseudoscalar component (that is
$g^P_{h_i{\tilde \chi^0_1}{\tilde \chi^0_1}} \neq 0$) and hence
significantly contribute to neutralino annihilation even at small
$v$. There is a kind of sum rule that relates the couplings
squared of the Higgses to neutralinos. Therefore, for the two
heavy eigenstates which are in general close in mass, we do not
expect  a large effect on the resulting relic density from Higgs
mixing alone. A noteworthy exception occurs when, for kinematical
reason, only one of the resonances is accessible to neutralino
annihilation. That is for example the case when
$m_{h_2}<2\mnt{1}\simeq m_{h_3}$.

\section{Results}

We now turn to the numerical analysis and present results for two
typical scenarios for which the relic density is in agreement
with WMAP: the mixed bino-Higgsino LSP that annihilates into gauge
bosons and the rapid annihilation through a Higgs resonance, the
so-called Higgs funnel. More comprehensive results can be found in
~\cite{Belanger:2006qa}. We always impose a constraint from the
eEDM, $d_e<2.2\times 10^{-27} {\rm e\,cm}$~\cite{Hagiwara:2002fs},
including one- and two-loop contributions. When
$\tan\beta$ is not too large and sfermions of the first and second
generations are heavy, this constraint usually forces
$\sin\phi_\mu\approx 10^{-2}$. In addition, the free parameter
$\phi_e$ is adjusted so that this constraint is  satisfied.

\subsection{The mixed bino-Higgsino LSP}
\label{sec:Higgsino}

\begin{figure}
  \includegraphics[height=.3\textheight]{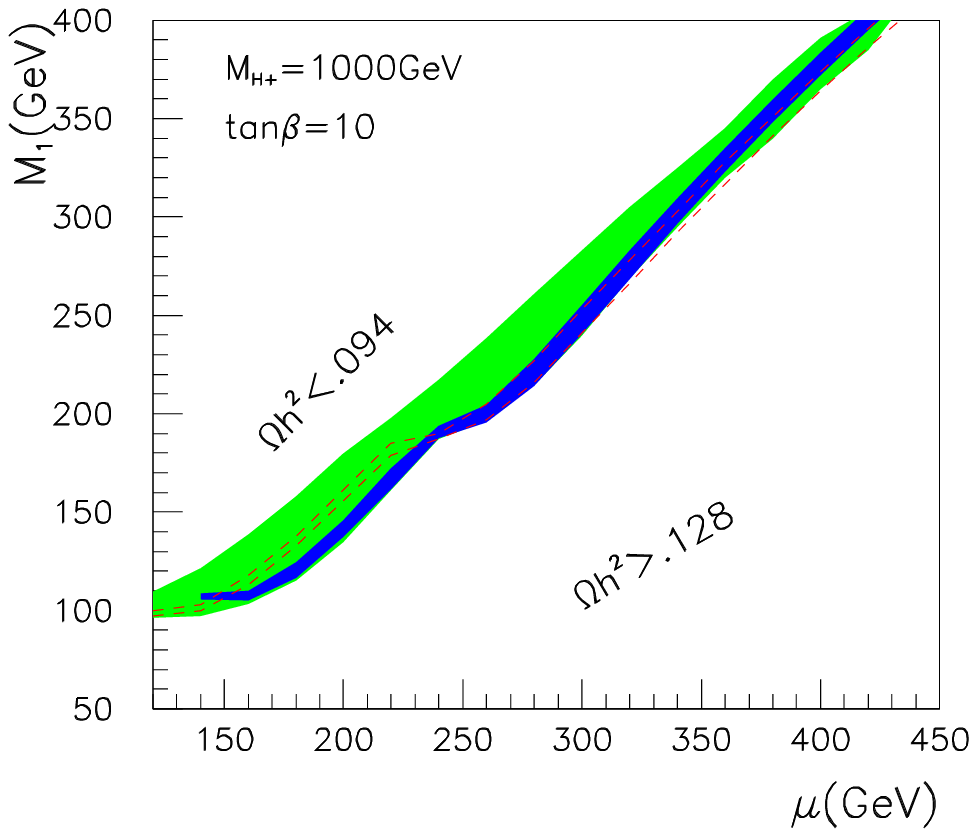}
  \includegraphics[height=.3\textheight]{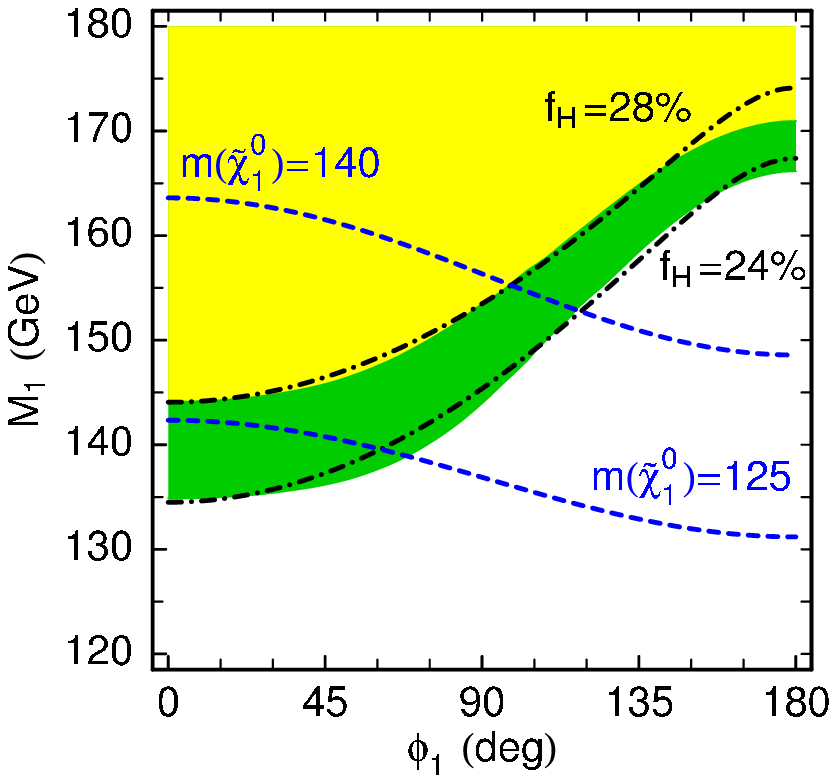}
  \caption{(a) The $2\sigma$ WMAP bands
  for  $\tan\beta=10$, $m_{H^+}=M_S=1$~TeV
  a) in the $M_1$--$\mu$
           plane for $A_t=1$~TeV and
           for all phases zero (blue),
           for $\phi_\mu=180^\circ$ (or $\mu<0$) and all other phases zero
           (dashed red lines)
           and for arbitrary phases (green). ,
           (b)
  $M_1$--$\phi_1$ plane for $\mu=200$~GeV,  $A_t=1.2$~TeV, $\phi_t=\phi_\mu=0$.
  In the yellow  region $\Omega h^2$ is below the
  WMAP bound, and in the white region it is too large.
  Superimposed are  contours of constant LSP mass (dashed lines) and contours of constant
  LSP Higgsino fraction (dash-dotted lines).}
\label{fig:m1mumh1000}
\end{figure}

When  all scalars except the light Higgs are heavy,
$M_S=m_{H^+}=1$~TeV, the most efficient annihilation of a pair of
neutralinos with a mass of the order of 100~GeV is through its
Higgsino component. In the real MSSM, one needs a Higgsino
admixture of roughly 25\%--30\% for the relic density to be within
the WMAP range \cite{Belanger:2004hk, Masiero:2004ft}. In terms of
fundamental MSSM parameters this means $M_1\approx \mu$.
Figure~\ref{fig:m1mumh1000}a  displays the allowed 2$\sigma$ WMAP
band in the $M_1-\mu$ plane. The main annihilation mechanisms then
are $\nt_1\nt_1\to WW/ZZ$  through $t$-channel chargino and
neutralino exchange, as well as $\nt_1\nt_1\to t\bar t$ when
kinematically allowed. The latter proceeds through $s$-channel $Z$
or $h_1$ exchange. The LSP Higgsino fraction determines the size
of the annihilation cross-section because it directly enters both
the $\nt_1\chpm_iW^\mp$ and $\nt_1\nt_j Z$ vertices.

When allowing all phases to vary arbitrarily, thus increasing the
number of free parameters, it is natural to expect a widening of
the allowed band, see the green (light grey) band in
Fig.~\ref{fig:m1mumh1000}a where we display models for which all
constraints are satisfied for at least one combination of phases.
Since the eEDM constraint strongly constrains $\phi_\mu$,  the
widening of the allowed band is mostly due to $\phi_1$.

\begin{figure}
  \includegraphics[height=.3\textheight]{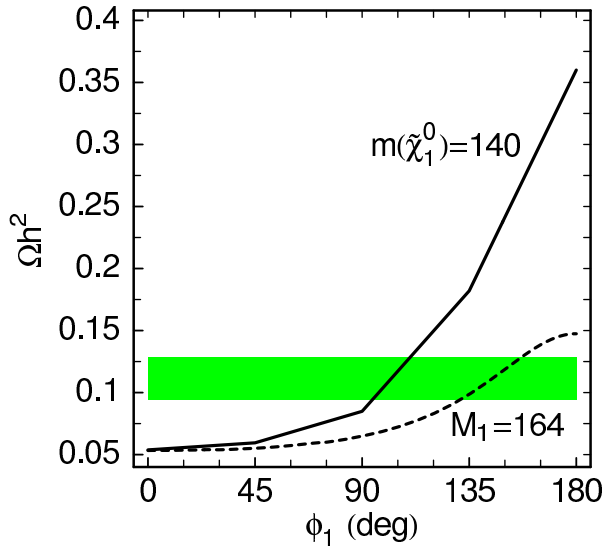}
 \caption{$\Omega h^2$ as a function of $\phi_1$ for the parameters
  of Fig.~\ref{fig:m1mumh1000}b; the dashed line is for fixed $M_1=164$~GeV,
  while for the full line $M_1$ is adjusted such that $\mnt{1}=140$~GeV.
  The green (grey) band shows the $2\sigma$ WMAP-allowed range.}
\label{fig:mlsp140}
\end{figure}

Figure~\ref{fig:m1mumh1000}b shows the WMAP band in the
$M_1$--$\phi_1$ plane  for $\mu=200$~GeV,
$\tan\beta=10, A_t=1$~TeV and all other phases set to zero. Also shown are contours of constant LSP
mass as well as contours of constant LSP Higgsino fraction $f_H$.
We can make several observations.
First, the mass of the LSP increases with $\phi_1$. On the one
hand this induces a decrease in the LSP pair-annihilation
cross-sections. On the other hand, since the chargino mass is
independent of $\phi_1$, the NLSP--LSP mass splitting is reduced,
making coannihilation processes with $\ch_1$ (and also $\nt_2$)
more important.
Second, the Higgsino fraction decreases with increasing $\phi_1$.
This modifies the  LSP couplings to gauge bosons and leads to a
decrease in the dominant $\nt_1\nt_1\to WW/ZZ$ cross-sections and
thus a higher value for the relic density. This phase dependence
in the couplings is predominantly what determines the shift in the
value of the relic density, to wit the  almost perfect match
between contours of constant $\Omega h^2$ and those of constant
$f_H$ in Fig.~\ref{fig:m1mumh1000}b. The deviation near
$\phi_1\sim 180^\circ$ comes from chargino/neutralino
coannihilations. To isolate the effect that comes solely from
modifications in the couplings, we display in Fig.~\ref{fig:mlsp140}
the variation of $\Omega h^2$ as function of $\phi_1$ for constant
LSP mass, as compared to the variation of $\Omega h^2$ for constant
$M_1$. The former is more pronounced reaching almost an order of
magnitude. In this scenario with a mixed bino-Higgsino LSP, the
dependence of masses and couplings on $\phi_1$ work against each
other, so that taking out the kinematic effects actually enhances
the variation of $\Omega h^2$.

\subsection{Annihilation through Higgs}

In the Higgs sector, nonzero phases  can induce
scalar-pseudoscalar mixing as well as important changes in the
masses. Moreover,
the scalar and pseudoscalar couplings of the Higgses to the LSP depend
on the phases in the neutralino sector. One can therefore expect large differences between the
real and complex MSSM in the Higgs-funnel region. Since the relic
density is very sensitive to the mass difference $\Delta
m_{{\tilde \chi^0_1} h_i}= m_{h_i}-2{m_{\tilde \chi^0_{1}}}$
~\cite{Allanach:2004xn}, it is important to
disentangle the phase effects in kinematics and in couplings. For
a case study of  the Higgs funnel, we choose
\begin{equation}
  M_1=150{\rm\;GeV},~\tan\beta=5,~
  M_S=500{\rm\;GeV},~ A_t=1200{\rm\;GeV},~ \mu=500{\rm\;GeV}.
\label{eq:funnelpar}
\end{equation}
leading to a dominantly bino LSP and  small  mixing in the Higgs
sector for $\phi_t\neq 0$.

\begin{figure}
\includegraphics[height=.3\textheight]{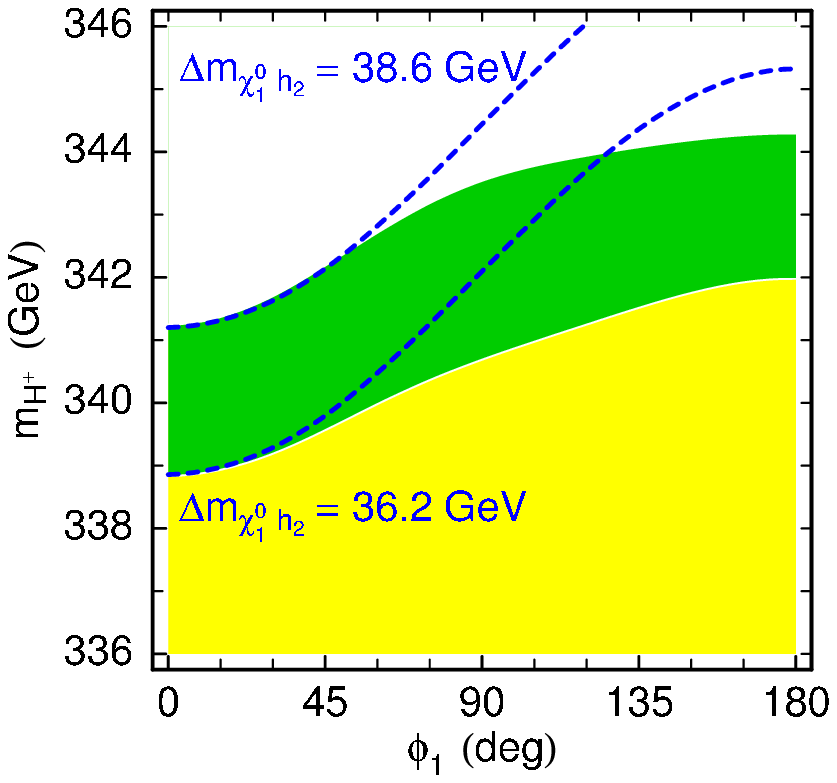}
 \includegraphics[height=.3\textheight]{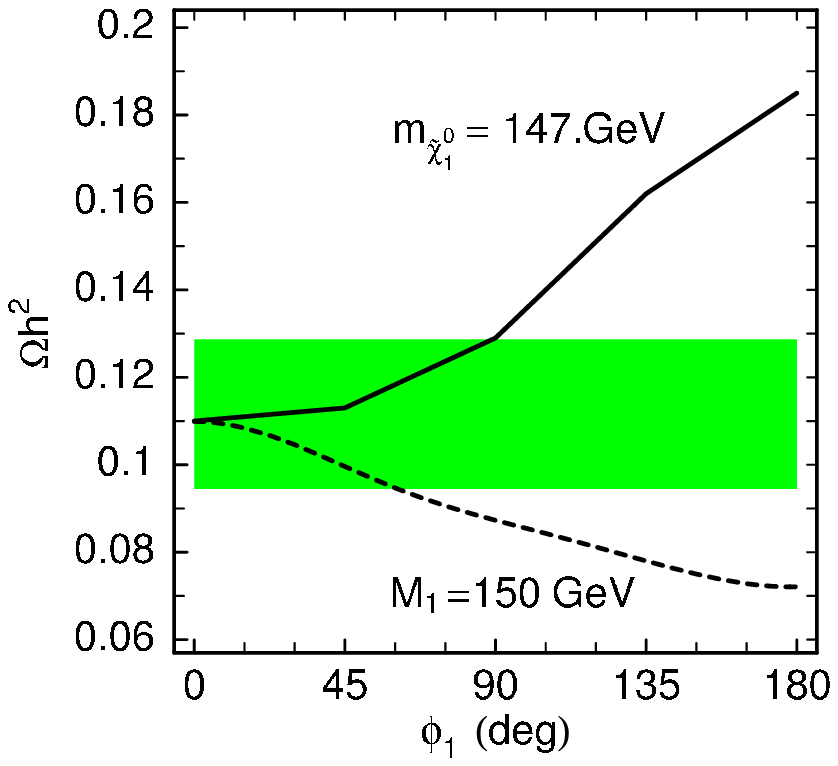}
\caption{a) The $2\sigma$ WMAP bands (green/dark grey) in the
  $m_{H^+}$--$\,\phi_1$ planes for
  the parameters of Eq.~(\ref{eq:funnelpar}) and
   $\phi_t=0$ .
  Contours of constant mass differences
  $\Delta m_{\nt_1h_2}=m_{h_2}-2\mnt{1}$ are also displayed.
  In the yellow  (light grey) region, $\Omega h^2$ is below the WMAP range.
  b) $\Omega h^2$ as a function of $\phi_1$ for
  $m_{H^+}=340$~GeV and the value of $M_1$ adjusted so that
  $\Delta m_{{\tilde \chi^0_1} h_2}$ stays constant (full line).
  For comparison, the variation of $\Omega h^2$ for fixed
  $M_1=150$~GeV is shown as a dashed line.}
\label{fig:at}
\end{figure}

In this scenario with relatively small $\mu$,  varying the phase
of $\phi_t$ does not have a large impact since the
scalar-pseudoscalar mixing never exceeds 8\%. The main effect of
$\phi_t$ on the value of the relic density can be explained by
shifts in the physical masses and position of the resonance.  On
the other hand, the phase $\phi_1$  changes the neutralino masses
and mixing, and hence also the neutralino--Higgs couplings,
Eq.~(\ref{eq:gs}). In  Fig.~\ref{fig:at}a we show the WMAP-allowed
regions in the $m_{H^+}$--$\phi_1$ plane for $\phi_t=0$. For a
given value of  $m_{H^+}$, when increasing $\phi_1$, the relic
density drops. This is because the mass of the neutralino
increases slowly, resulting in a smaller $\Delta m_{{\tilde
\chi^0_1} h_2}$. Adjusting $\mnt{1}$ or $m_{h_2}$ such that the
mass difference stays constant, we find rather that the relic
density increases with $\phi_1$, Fig.~\ref{fig:at}b. The maximum
deviation which comes purely from modifications in the couplings
can reach 70\%. This can be readily understood from the phase
dependence of the couplings of $h_{2,3}$ to the LSP. For
$\phi_1=0$, the coupling of $h_2$ is predominantly pseudoscalar
and thus gives the dominant contribution to neutralino
annihilation.  For $\phi_1=90^\circ$, the coupling of $h_3$  has a
large pseudoscalar component and also contributes  to the
annihilation. When
increasing $\phi_1$ further (up to $180^\circ$), $h_2$ exchange
again dominates, however with a  coupling to neutralinos smaller
than for $\phi_1=0$. Thus one needs a smaller mass splitting
$\Delta m_{{\tilde \chi^0_1} h_2}$ for $\Omega h^2$ to fall within
the WMAP range, see Fig.~\ref{fig:at}a.

When the mixing is large in the Higgs sector, say for $\mu=1$TeV, the
phase  $\phi_t$ affects the scalar-pseudoscalar mixing as well as
the mass splitting between $h_2,h_3$. The latter can reach several
GeV's. This means that $\phi_t$ can induce very large shifts in
the relic density which very sensitively depends on the mass
difference between the LSP and the predominantly pseudoscalar
Higgs. Isolating the phase dependence of $\Omega h^2$ due to the
scalar-pseudoscalar mixing by keeping $\Delta m_{{\tilde \chi^0_1}
h_3}$ constant, we found an increase in $\Omega h^2$ relative to
the $\phi_t=0$ case by almost an order of magnitude~\cite{Belanger:2006qa}. This is
however far less than the huge shifts of several orders of
magnitude found in Ref.~\cite{Gondolo:1999gu} for fixed values of
$m_{H^+}$ when a Higgs pole is passed.

\section{Conclusions}

We have shown here that in the CPV-MSSM, the relic density could
be quite different as compared to that in the MSSM with vanishing
phases, the variations in $\Omega h^2$ often exceeding the $\sim 10\%$ range
 of the WMAP bound. In many cases, a large part of the
phase dependence can be explained by changes in the masses of the
involved particles.  However, in some cases disentangling the kinematic
effects also leads to an enhancement of the phase dependence.
 This happens, for instance, in the
case of a mixed bino-Higgsino LSP, where we have found effects of
almost an order of magnitude from modifications in the couplings
due to nonzero CP phases.

 When aiming at a precise
prediction of the neutralino relic density from collider
measurements, it is clear that one does not only need precise
sparticle spectroscopy
--- one also has to precisely measure the relevant couplings and
this certainly includes the determination of possible CP phases.
Whether parameters of the CPV-MSSM can be determined with
sufficient precision at the LHC or at a future linear collider  requires
careful investigation.

\section*{Acknowledgements}
\addcontentsline{toc}{section}{Acknowledgements}

This work was supported in part by GDRI-ACPP of CNRS and by grants
from the Russian Federal Agency for Science, NS-1685.2003.2 and
RFBR-04-02-17448. S.K. is supported  by an APART (Austrian
Programme for Advanced Research and Technology) grant of the
Austrian Academy of Sciences.



\bibliographystyle{aipproc}   

\end{document}